\documentstyle[eqsecnum,aps,psfig,epsf]{revtex}

\begin{document}




\title{Exact Diagonalization of the Hamiltonian for\\Trapped Interacting
 Bosons in Lower Dimensions}


\author{Tor Haugset and H\aa rek Haugerud}

\address{Department of Physics, University of Oslo,\\
P.O. Box 1048 Blindern, N-0316 Oslo, Norway}


\bibliographystyle{unsrt}

\maketitle

 
\begin{abstract}
We consider systems of a small number of interacting bosons confined to
harmonic potentials in one and two dimensions. By exact numerical
diagonalization of the many-body Hamiltonian we determine the low lying 
excitation energies and the ground state energy and density profile. We 
discuss the dependence of these quantities on both interaction strength $g$ 
and particle number $N$. The ground state properties are compared to the 
predictions of the Gross-Pitaevskii equation, and the agreement is
surprisingly good even for relatively low particle numbers. 
We also calculate the specific heat based on the obtained energy spectra.
\end{abstract}




\section{Introduction}
\label{section:intro}
Recent observations of Bose-Einstein condensation in alkali gases confined to 
harmonic potentials \cite{sod,AEMWC} has led to an extensive study of the 
ground state and low energy properties of trapped interacting Bose gases. A 
major part of this work, see e.g. \cite{BP,kanwal,EDCRB,RHBE,ERBDC,DGGPS,esry},
 presents calculations based on the Bogoliubov approximation or Hartree-Fock 
theory. Predicted excitation frequencies agree well with experiments 
\cite{JEMWC,MADKDTK}. The energy spectrum has also been described in terms
of a hydrodynamic formalism \cite{stringari}, which agrees with Bogoliubov 
theory at low energy \cite{DGGPS,FR}. 
In the Bogoliubov approximation the
system is described in terms of a dominating condensate. Fluctuations induced
by finite temperature and interactions are assumed small. At zero temperature,
when the effects of excited particles are neglected, the condensate satisfies
the Gross-Pitaevskii equation \cite{gross,pitaevskii,griffin}. 

The Bogoliubov approximation is applicable for systems with a large number of
particles. In this article we will, on the contrary, consider systems
 with relatively few particles, i.e. $N\sim 10-40$. Finite $N$ effects are 
then important, and one should in principle use the complete machinery of 
many-body theory when describing these systems. Here, we present the results 
of such calculations, limiting ourselves to one and two dimensions.
The use of such 
time- and computer memory consuming methods puts severe limitations on the size
 of the system one may describe. On the other hand, it is for such systems the 
corrections to mean-field calculations are important. 

The results presented here are therefore not immediately relevant for the 
experiments performed with trapped Bose gases so far. Our aim is rather to 
gain some insight into the effects of the approximations which form the basis 
of the mean-field theories mentioned above, in the limit of small 
particle numbers. We are also interested in effects which are specific to 
one and two dimensions. In addition, future experiments in 
lower dimensions may explore the range of parameters used here.

The plan of the article is as follows. In the next section we recall the
many-body formalism and describe the steps leading to the diagonalization of
the many-body Hamiltonian. The Gross-Pitaevskii approximation, to which we 
compare our results for the ground state, is discussed in Section 
\ref{section:mf}. Numerical 
results are presented and discussed in Section \ref{section:results}. Finally 
we draw some conclusions in the last section. The appendix discusses 
effective couplings in highly anisotropic systems.

\section{Many-body description}
\label{section:mp}
As a model for a trapped interacting Bose gas we study the Hamiltonian density
\begin{eqnarray}
{\cal H} = \hat{\Psi}^{\dagger}({\bf r})\left[-\frac{\nabla^2}{2m} + 
V_{ex}({\bf r})\right]\hat{\Psi}({\bf r}) + {\cal H}_I(\hat{\Psi}^{\dagger},
\hat{\Psi}),
\end{eqnarray}
where $\hat{\Psi}$ is a bosonic field operator and $V_{ex}$ is the external 
trapping potential. Units where $\hbar=1$ are used. 
We will only consider isotropic harmonic potentials of the
 form $V_{ex}({\bf r})=\frac{1}{2}m\omega^2r^2$. ${\cal H}_I$ describes the 
interaction. Assuming short-range two-body interaction, this 
may be written
\begin{eqnarray}
{\cal H}_I = g\hat{\Psi}^{\dagger}\hat{\Psi}^{\dagger}
\hat{\Psi}\hat{\Psi}({\bf r}).
                                                       \label{eq:Hi}
\end{eqnarray}
In the s-wave approximation the coupling constant, or interaction
strength $g$ is related to the 
scattering length $a$ by the equation $g = 2\pi a/m$ 
in three dimensions. In Appendix \ref{section:effcoupl} we discuss 
modifications of this relation in highly anisotropic harmonic traps.\\\\
{\em Harmonic modes.}
As a step towards finding the energy eigenvalues, we expand the field 
operator in a complete set of modes
\begin{eqnarray}
\hat{\Psi}({\bf r}) = \sum_k\varphi_k({\bf r})a_k.
\end{eqnarray}
Here $a_k$ annihilates a particle in the state $k$ and $a_k^{\dagger}$ creates
a particle in the same state. The 
operators satisfy the commutation relation $[a_k,a_l^{\dagger}] = 
\delta_{k,l}$, with other commutators vanishing. In translation invariant 
systems it is convenient to choose a set of plane waves. Here we choose to use
the harmonic oscillator eigenfunctions. The Hamiltonian then takes the form
\begin{eqnarray}
H \equiv \int d{\bf r} {\cal H}
= \sum_k\omega_ka_k^{\dagger}a_k
+ g \sum_{k,l,m,n}f_{klmn}a_k^{\dagger}a_l^{\dagger}a_ma_n.
                                               \label{exact-eq:Hamiltonian}
\end{eqnarray}
Here, $\omega_k$ is the single particle energy of level $k$, 
and $f_{klmn}$ is an overlap integral over four oscillator eigenfunctions
\begin{eqnarray}
f_{klmn} = \int d{\bf r}\varphi^*_k\varphi^*_l\varphi_m\varphi_n({\bf r}).
\end{eqnarray}
Notice that $f_{klmn}$ is symmetric under permutation of the indices.
For the complete set of eigenfunctions $\{\varphi_k\}$ we will choose Hermite 
or Laguerre polynomials for one and two dimensions, respectively. The 
corresponding overlap integrals can be calculated numerically, though 
some of them are tabulated \cite{GR}. The interaction conserves parity. In 
one dimension this leads to the constraint $k+l+m+n=$ even integer for 
non-zero $f$. In two dimensions the rotational symmetry of the external
potential
suggests the use of radial and angular quantum numbers. The interaction 
conserves angular momentum, and this again puts a constraint on the possible 
combinations of creation and annihilation operators. \\\\
{\em Many particle basis and diagonalization.} 
We apply the configuration interaction approximation and diagonalize the
Hamiltonian in a many particle basis, which in the occupation-number
representation may be written in the form
\begin{eqnarray}
|\psi_{\alpha}\!\rangle = |n_0n_1n_2\cdots n_K\!\rangle_{\alpha},
\end{eqnarray}
with $\alpha$ labeling the different distributions of particles.
Normalization to unity is assumed. Here $n_k$ is the particle number in the 
single particle state $k$. The particle number is conserved: $\sum_kn_k = N$.
In a practical calculation we 
must truncate the basis set at some upper state, here denoted by $K$. This 
sets an upper limit for the single particle energy of the states considered. 
Such a truncation is however somewhat unnatural as far as the many particle 
energy is concerned, since it includes the many particle state 
$|0_0\cdots N_K\rangle$ with energy $N\omega_K$, but not the state
$|(N\!-\!1)_0\cdots 1_{K+1}\rangle$ which only has 
energy $\omega_{K+1}+(N\!-\!1)\omega_0$. A more consistent way of truncating 
the basis set is therefore to include only those states which have a total 
energy up to some maximal value $E_{max}$. In the actual calculations, 
$E_{max}$ will be raised in steps until convergence is reached. 
The creation operator and
annihilation operator act as follows
\begin{eqnarray}
a_k|n_0\cdots n_K\!\rangle &=& \sqrt{n_k}|n_0\cdots n_k-1\cdots n_K\!\rangle,
\nonumber\\
a_k^{\dagger}|n_0\cdots n_K\!\rangle &=& \sqrt{n_k+1}|n_0\cdots n_k+1\cdots n_K
\!\rangle.
\end{eqnarray}
The number operator is thus $a_k^{\dagger}a_k$ with eigenvalue $n_k$. 

With a basis set at hand, we may calculate the Hamiltonian matrix elements
\begin{eqnarray}
H_{\alpha',\alpha} = \langle\!\psi_{\alpha'}|H|\psi_\alpha\!\rangle = 
H_{0\alpha}\delta_{\alpha',\alpha} + H_{I\alpha',\alpha}.
\end{eqnarray} 
The free part of the Hamiltonian contributes to the diagonal terms with an 
amount $\sum_k\omega_kn_k$, while the interaction part yields the contribution 
\begin{eqnarray} 
g\sum_{k,l,m,n}f_{klmn}\langle\!\psi_{\alpha'}|a_k^{\dagger}a_l^{\dagger}a_m
a_n|\psi_\alpha\!\rangle.
\end{eqnarray}
The symmetries of the interaction can be used to reduce the Hamiltonian matrix 
to a block diagonal form. Splitting the complete Fock space into several 
symmetry invariant subspaces reduces the dimensions of the Hamiltonian matrix
considerably and enables numerical diagonalization of much larger 
systems. In one dimension there is an even 
and an odd parity subspace and the ground state is a superposition of even 
parity states. In two dimensions the subspaces are labeled by the total 
angular momentum of the many particle state and the ground state has zero 
angular momentum.  

The Hamiltonian matrix is relatively sparse and the iterative Lanczos method 
\cite{LANCZOS} is thus well suited for finding the ground state and 
the lowest excited states. This method enables us to diagonalize much larger 
matrices than would be possible with standard library routines. However, 
we use the latter when finding the complete set of energy levels needed for 
calculating thermodynamic quantities. 

We would here like to emphasize that we have checked 
the programs thoroughly: They are seen to reproduce well known results
in the case of zero coupling. We have also done some calculations
for smaller systems by hand, and find that both the generation of the
Hamiltonian matrix and the diagonalization works correctly.\\\\
{\em Density profile.} Once the Hamiltonian is diagonalized, we have the 
coefficients for the low lying states, and especially for the ground state. 
This enables us to find the ground state density distribution 
$\rho({\bf r})$. The density operator may be written
\begin{eqnarray}
\hat{n}_{\psi} = \sum_{k,l}|k\rangle n_{kl}\langle l|,
\end{eqnarray}
where $|k\rangle$ is a single particle state, and $n_{kl} = \langle
\psi|a_l^{\dagger}a_k|\psi\rangle$. The density is then $\rho({\bf r}) = 
\langle{\bf r}|\hat{n}_{\psi}|{\bf r}\rangle$. For a state $|\psi\rangle=
\sum_{\alpha}C_{\alpha}|\psi_{\alpha}\!\rangle$ this reads
\begin{eqnarray}
\rho({\bf r}) = \sum_{k,l}\varphi^*_l\varphi_k({\bf r})\sum_{\alpha,\alpha'}
C^*_{\alpha'}C_{\alpha}\langle\!\psi_{\alpha'}|a_l^{\dagger}a_k|\psi_{\alpha}
\!\rangle.
\end{eqnarray}\\\\
{\em Dipole mode.} The excitations in this system will in general have 
interaction dependent energies. However, as discussed by Fetter and Rokhsar 
\cite{FR}, there exists a dipole mode for each spatial dimension of the trap, 
which corresponds to a harmonic oscillation of the center of mass of the 
condensate.
In \cite{FR} the corresponding raising operator was constructed in the first 
quantized formalism:
\begin{eqnarray}
A^{\dagger}_{\beta} = \sum_{i=1}^{N}b_{\beta i}^{\dagger},
\end{eqnarray}
where $b_{\beta i}^{\dagger}$ is the raising operator for particle number $i$ 
in dimension $\beta$. (We use a different notation than in \cite{FR}, in 
order to avoid confusion with quantities defined here.) With 
a two-body potential of the form $V({\bf r}_i\!-\!{\bf r}_j)$ 
it is easily shown
that the excitation energy is $\omega$, i.e. independent of the interaction. 
Thus, for each energy level there must be a ladder of levels with energy 
spacing $\omega$ above. These dipole modes have been found in calculations 
based on the Bogoliubov approximation \cite{kanwal,ERBDC} and in the 
hydrodynamic description \cite{stringari}. The occurrence of these states will 
serve as a check of convergence of the calculated energy levels. For 
completeness we note that the raising operator in terms of creation and 
annihilation operators in one dimension takes the form
\begin{eqnarray}
A^{\dagger} = \sum_k\sqrt{k+1}a_{k+1}^{\dagger}a_k.
\end{eqnarray}
Here $k$ labels the Hermite polynomials. Commuting the raising operator with 
the free particle Hamiltonian we find
\begin{eqnarray}
[H_0,A^{\dagger}] = \omega A^{\dagger},
\end{eqnarray}
and the excitation energy is thus $\omega$. The vanishing of the commutator 
of $A^{\dagger}$ with the interaction term is less obvious in this formalism. 
The commutator becomes
\begin{eqnarray}
[H_I,A^{\dagger}] = 2g\!\!\sum_{k,l,m,n}a^{\dagger}_k\sqrt{l+1}(f_{kmnl+1}
a^{\dagger}_na_l
- f_{kmnl}a^{\dagger}_{l+1}a_n)a_m,
\end{eqnarray}
and for each combination $a^{\dagger}_ka^{\dagger}_la_ma_n$ the prefactor 
vanishes. This is due to the following relation between the overlap integrals:
\begin{eqnarray}
\sqrt{d+1}f_{abcd+1} + \sqrt{c+1}f_{abc+1d}
= \sqrt{a}f_{a-1bcd} + \sqrt{b}f_{ab-1cd}.
                                                  \label{exact-eq:frelation}
\end{eqnarray}\\\\
{\em Breathing mode.}  In two dimensions the many-body Hamiltonian
has an $SO(2,1)$-symmetry discovered by Pitaevskii and Rosch \cite{PR}.
This symmetry gives rise to excitations of energy $2\,\omega$ identified with
the breathing mode of the condensate. The excitation spectra for the planar
systems must therefore contain ladders both with energy spacing $\omega$ and
$2\,\omega$. The symmetry is broken when a UV energy cut-off is introduced
in the system. Such a cut-off is inherent in our calculation. We therefore
expect that the level spacing $2\,\omega$ will be less accurate than the
spacing $\omega$ caused by the dipole excitations.\\\\

\section{Mean field Approximation}
\label{section:mf}
One of the purposes of this paper is to compare the predictions of the 
many-body calculation with those of mean field theory for quantities such as 
the ground state energy and density distribution. The condensate is in the 
Gross-Pitaevskii (GP) approximation \cite{gross,pitaevskii} described by a 
classical, macroscopic Bose field governed by the Hamiltonian density per 
particle
\begin{eqnarray}
{\cal H}_{cl}/N = \Phi^*({\bf r})\left(-\frac{\nabla^2}{2m} + V_{ex}
({\bf r})\right)\Phi({\bf r})
+ gN\left(\Phi^*\Phi({\bf r})\right)^2.
\end{eqnarray}
The field is here rescaled to satisfy $\int d{\bf r} \Phi^*\Phi = 1$, thus 
the factor $N$ in front of the interaction term. Notice that the coupling $g$ 
and the particle number $N$ occur only in the combination $gN$, as opposed 
to the many-body description. Minimalization of the corresponding Hamiltonian 
$H_{cl}=\int d{\bf r}{\cal H}_{cl}$ leads to the non-linear Schr\"{o}dinger 
(or Gross-Pitaevskii) equation
\begin{eqnarray}
\left(-\frac{\nabla^2}{2m} + V_{ex}({\bf r}) - \mu\right)\Phi({\bf r}) + 
2gN\Phi^*\Phi\Phi({\bf r}) = 0.
                                                          \label{GPequation}
\end{eqnarray}
The chemical potential introduced is given by the normalization of $\Phi$. 
This equation can be derived from the Hartree-Fock-Bogoliubov approximation 
when the effect of excited particles is neglected \cite{griffin}.\\\\
{\em Oscillator basis.} The equation for $\Phi$ may be solved 
introducing a harmonic oscillator basis
\begin{eqnarray}
\Phi({\bf r}) = \sum_kc_k\phi_k({\bf r}).
\end{eqnarray}
Normalization now corresponds to $\sum_k|c_k|^2 = 1$.
The equation then takes the form
\begin{eqnarray}
\sum_k(\omega_k - \mu)c_k\phi_k({\bf r})
+ 2gN\sum_{k,l,m}c^*_kc_lc_m\phi^*_k\phi_l\phi_m({\bf r}) = 0.
\end{eqnarray}
Multiplying with $\phi^*_n$ and integrating we finally end up with the set 
of  equations
\begin{eqnarray}
(\omega_n - \mu)c_n + 2gN\sum_{k,l,m}c^*_kc_lc_mf_{klmn} = 0,
\end{eqnarray}
where $f$ again is the overlap integral over four harmonic oscillator
eigenfunctions. The basis set is truncated at some level $K$, and $K$ 
is raised 
until convergence is reached. The corresponding finite set of equations may 
be solved by use of the Newton-Raphson method. This method was applied 
in \cite{ERBDC} for the case of a three-dimensional Bose gas. Solutions in 
one dimension have also been obtained earlier \cite{RHBE}. Again the symmetry
of the ground state may be used to choose the proper 
subset of eigenfunctions. In one dimension we only have to consider even 
functions. In two dimensions only zero angular momentum eigenfunctions 
contribute. 

Once the coefficients $c_k$ are determined one must check if the 
normalization condition is satisfied. If not, a different chemical potential 
is chosen, and the process is repeated until normalization is obtained. 
Having the correct coefficients, we easily find the ground state density 
profile
\begin{eqnarray}
\rho({\bf r}) = |\sum_kc_k\phi_k({\bf r})|^2.
\end{eqnarray}
In addition the ground state energy per particle reads
\begin{eqnarray}
\sum_k\omega_k|c_k|^2 + gN\sum_{klmn}c^*_kc^*_lc_mc_nf_{klmn}.
\end{eqnarray}

\section{Numerical results}
\label{section:results}
In the following we present the results of the numerical diagonalization of
the many-body 
Hamiltonian. In one dimension we have truncated the many-body basis at cut-off
energy $E_{max} = 38\,\omega$. This corresponds to a basis with blocks of up 
to 80524 many-body states. In two dimensions the number of overlap 
integrals for
a given $E_{max}$ is considerably higher than in one dimension. Moreover,
the generation of matrix elements turns out to be very time consuming.
We have therefore stopped at $E_{max} = 14\,\omega$ with a basis with blocks 
of up to 4532 states in two dimensions. Energies are measured in units of 
the oscillator
frequency and lengths in units of the typical oscillator width 
$1/\sqrt{m\omega}$. $g$ refers to the corresponding dimensionless interaction
strength $(\tilde{g}_{1D},\tilde{g}_{2D})$ as obtained in the appendix.

\begin{figure}[htb]
\begin{center}
\mbox{\psfig{figure=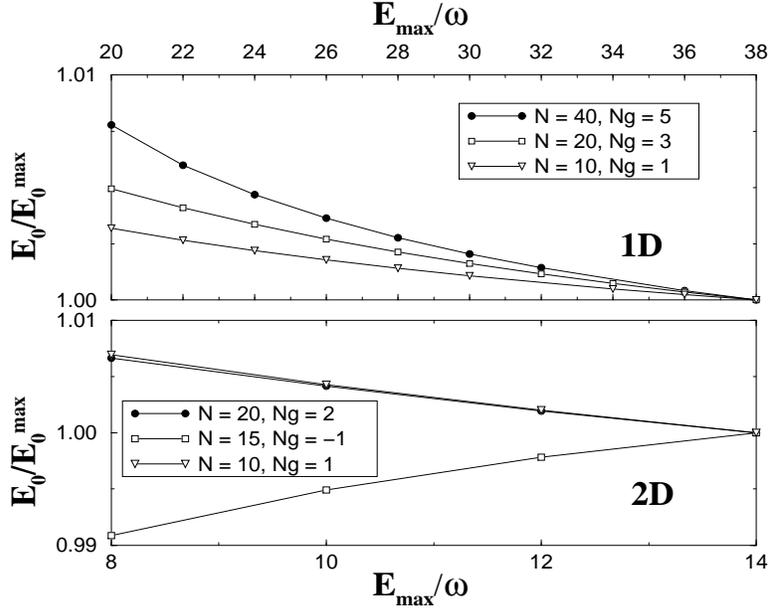,width=10cm,angle=270,height=8cm}}
\end{center}
\caption{\footnotesize Convergence of the ground state energy. We plot the
ratio of the ground state energy $E_0$ for the given cut-off $E_{max}$
and the energy $E_0^{max}$ found using the highest $E_{max}$. 
The upper graph is for one dimension and the lower for two dimensions.}
\label{convergence}
\end{figure}

In Fig.\ 1 we show the ground state energy as function of the 
cut-off energy $E_{max}$. The plotted quantity is the ratio
of the ground state energy for the given $E_{max}$ and the energy found
using the highest $E_{max}$ considered. The upper graph is for one 
dimension and
the lower for two dimensions. In one dimension, the relative difference in
ground state energy going from $36\,\omega$ to $38\,\omega$ is seen to be about
$0.05\%$ or less. Going from $12\,\omega$ to $14\,\omega$ in two dimensions, 
the shift is about $0.2\%$. 
The convergence of the excitation energies is found to be considerably faster
for the lowest levels, but is slower at high energy. This will be seen below 
in the deviation of the interaction independent energies from the exact values
$n\omega$. 

\begin{figure}[htb]
\begin{center}
\mbox{\psfig{figure=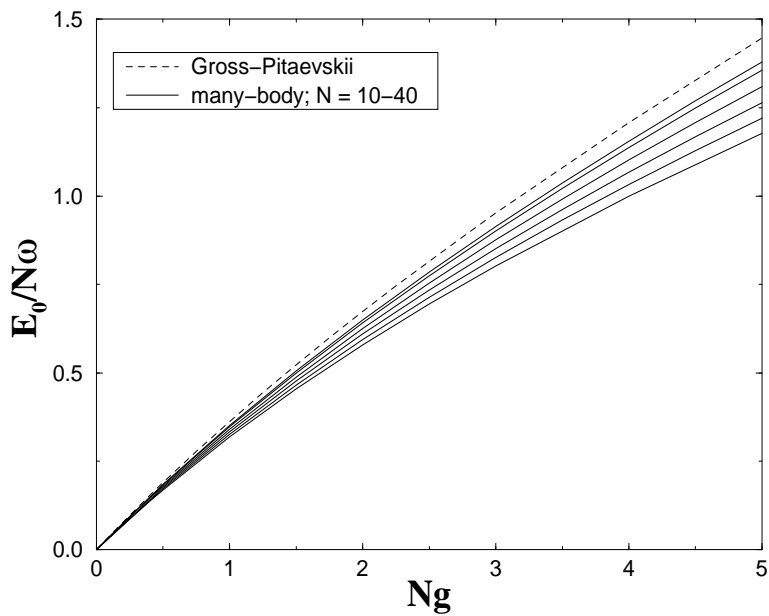,width=10cm,angle=270,height=8cm}}
\end{center}
\caption{\footnotesize Ground state energy $E_0$ per particle in 
one dimension measured in units 
of $\omega$ as function of interaction strength $g$
for particle numbers $N=10,12,15,20,30$ and 40 starting from below. 
The results of the many-body calculation converge towards the mean field
result as the particle number increases.}
\label{EoD1}
\end{figure}

\subsection{Ground state energy}

In this subsection we show the effect of the interaction on the energy of the 
ground state. In all figures the ideal gas ground state energy is set to 
zero.\\\\
{\em Linear Bose gas.} Fig.\ 2 shows the ground state energy in one dimension,
both in the many-body description 
and in the Gross-Pitaevskii approximation.
For a given $Ng$ the energy resulting from the many-body calculation is seen to
converge towards the mean field value as the particle number $N$
is increased. The discrepancy between the approximations increases with 
$Ng$.

\begin{figure}[htb]
\begin{center}
\mbox{\psfig{figure=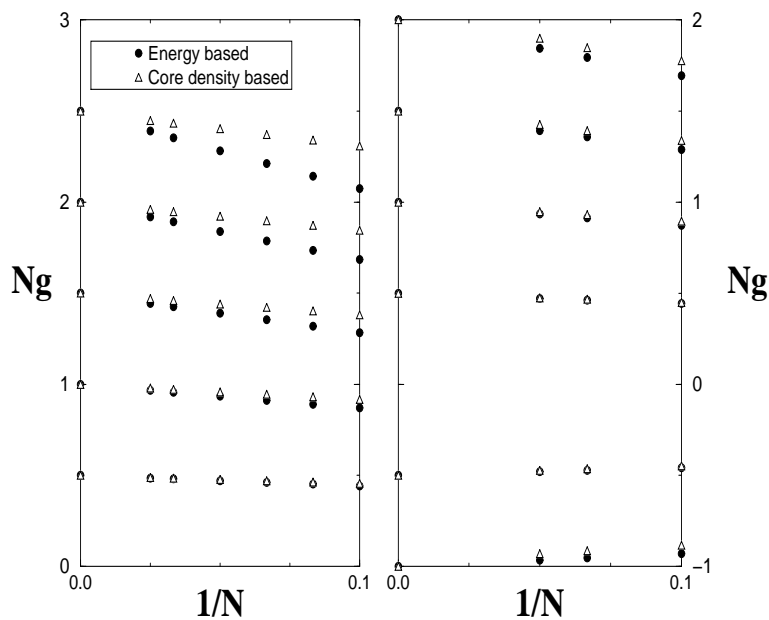,width=10cm,angle=270,height=8cm}}
\end{center}
\caption{\footnotesize Effective coupling $Ng$ in one (left) and two (right)
dimensions. The corresponding values of $Ng_0$ are found at $1/N=0$.
Effective couplings based on the energy and the core density of
the ground state are compared.}
\label{EffCoupl}
\end{figure}

The discrepancies found between the two approximations might
to some extent be removed by the introduction of an effective coupling constant
in the mean field theory. 
For a set of couplings $g_0$ and particle numbers $N$ we find the 
rescaled coupling which in the GP-approximation yields the corresponding
energy given by the 
many-body calculation. In Fig.\ 3 this rescaled coupling multiplied
by $N$ is plotted (filled circles) 
as a function of $1/N$ for various values of the bare coupling. The 
corresponding value of $Ng_0$ 
is found at $1/N=0$. For the range of parameters considered, 
the lines look very linear. Below, we compare these results with the
effective coupling based on calculations of the core density in the
ground state.\\\\
{\em Planar Bose gas.} The insertion in Fig.\ 4 shows the
Gross-Pitaevskii ground state energy in two 
dimensions as function of $Ng$. The energy falls rapidly with decreasing, 
negative coupling. With a basis of up to 28 oscillator 
eigenfunctions, we have not been able to reach convergence near 
$E_0/N=-\omega$.
It seems though that $dE_0/d(Ng)$  approaches infinity as we get 
closer to this point, thus indicating that no stable ground state may be 
formed for stronger attractive couplings. This is agreement with Pitaevskii's 
analysis of the system \cite{pitaevskii2D}, where it is shown that an 
attractive interaction leads to collapse of the gas for energies more than 
$\omega$ below the ideal gas ground state energy. Similar conclusions are 
drawn in \cite{WMM}.

\begin{figure}[htb]
\begin{center}
\mbox{\psfig{figure=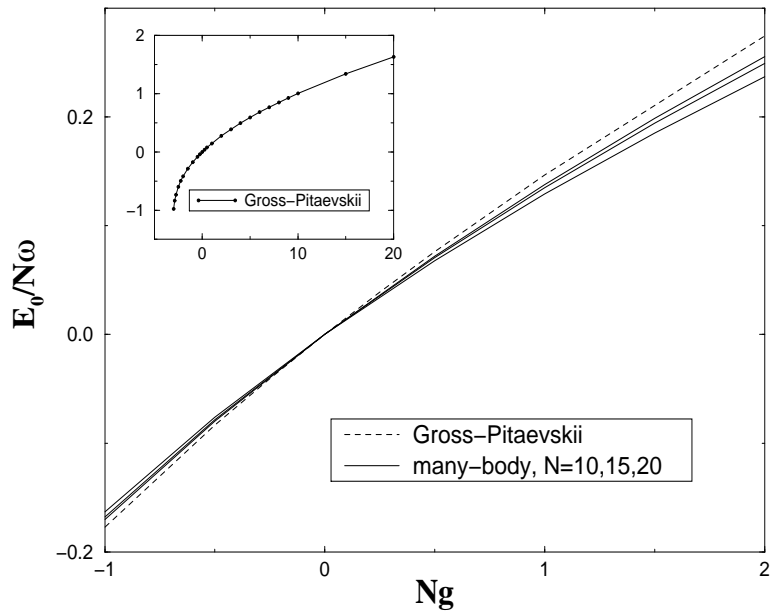,width=10cm,angle=270,height=8cm}}
\end{center}
\caption{\footnotesize Ground state energy $E_0$ per particle in two dimensions 
measured in units of $\omega$ as function of interaction strength $g$
for particle numbers $N=10,15$ and 20 starting from below. The results of the
many-body calculation are compared to the mean field
approximation (Gross-Pitaevskii), 
which is approached as the particle number increases.
The insertion shows the energy in the mean field
approximation over a broader range of the interaction strength.}
\label{EoD2}
\end{figure}

The Gross-Pitaevskii ground state energy is in Fig.\ 4 also plotted 
together with the results of
the many-body calculation for $N=10,15$ and 20. The mean field
approximation is approached as $N$ increases. 
We have calculated the
effective coupling $Ng$ which, used in the GP-approximation
reproduces the energy obtained in the many-body calculation. This effective
coupling is plotted as filled circles
in the right graph of Fig.\ 3. The results
are all for the highest cut-off energy $E_{max}=14\omega$. We again find
that $Ng$ depends nearly linearly on $1/N$. 

\begin{figure}[htb]
\begin{center}
\mbox{\psfig{figure=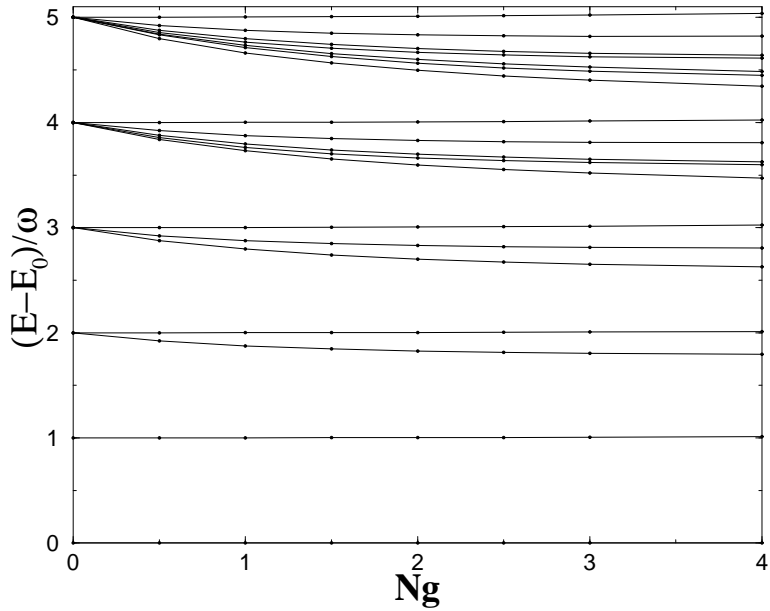,width=10cm,angle=270,height=8cm}}
\end{center}
\caption{\footnotesize Excitation energies $(E-E_0)/\omega$ in a 
one-dimensional system of
30 bosons as function of the interaction strength $g$.}
\label{dED1}
\end{figure}

\subsection{Excitations}

In Section \ref{section:mp} we saw that the excitation energies must consist 
of ladders with energy spacing $\omega$. This is due to the existence of the 
collective raising operator $A^{\dagger}$. In two dimensions the breathing mode
leads to additional ladders with spacing $2\,\omega$.
The bottom of each of the ladders 
must however be determined by other means, and here we present the
results of the numerical diagonalization of the many-body Hamiltonian.\\\\
\begin{figure}[htb]
\begin{center}
\mbox{\psfig{figure=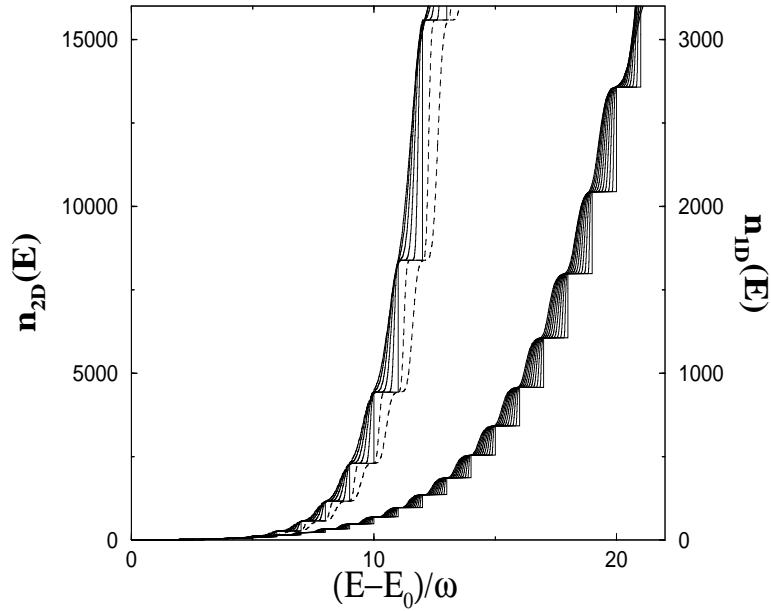,width=10cm,angle=270,height=8cm}}
\end{center}
\caption{\footnotesize Summed density of states in a one-dimensional (right curves, 
$n_{1D}$)
and two-dimensional (left curves, $n_{2D}$) system of 20 and 15 
bosons respectively. The
interaction strength is in the range $Ng=0,0.2,\ldots,2$ in one dimension and
$Ng=-1,-0.5,\ldots,2.5$ in two dimensions. $Ng$ increases from right to left,
and negative interaction strengths are indicated by dashed lines.}
\label{sDOS}
\end{figure}
{\em Linear Bose gas.} In Fig.\ 5 the lowest excitation energies in one 
dimension are given as function of $Ng$ for the case of 30 particles. The 
degeneracy at zero coupling is split by the interaction. For each level in 
the ideal case, one level remains independent of the coupling. These levels 
correspond to excitations described by $A^{\dagger}$ from the ground state. 
The others spread out below, and above each of these we find the 
mentioned ladder of levels. Similar results are obtained for $N=10,12,15,20$ 
and 40. The relative shift in the energies going from $N=10$ to $N=40$ with
a given $Ng$ is less than $1\%$ for the range of parameters presented in 
Fig.\ 5.

The splitting of the degeneracy by the interaction leads to a more even 
distribution of energy levels than in the free case. As an illustration of 
this, Fig.\ 6 shows the number of energy levels $n_{1D}(E)$ (right curves) 
below a given energy 
$E$ measured from the ground state. We have taken $N=20$, and the 
coupling ranges from $Ng=0-2$ in steps of 0.2. This summed density of states 
is indeed smoothed by the presence of the interaction. Note that the curves 
coincide at points $(E\!-\!E_0)/\omega = n$. This happens because all 
interaction-dependent excitation energies lie in the gap below the 
corresponding undisturbed level for the considered values of the coupling.\\\\ 
\begin{figure}[htb]
\begin{center}
\mbox{\psfig{figure=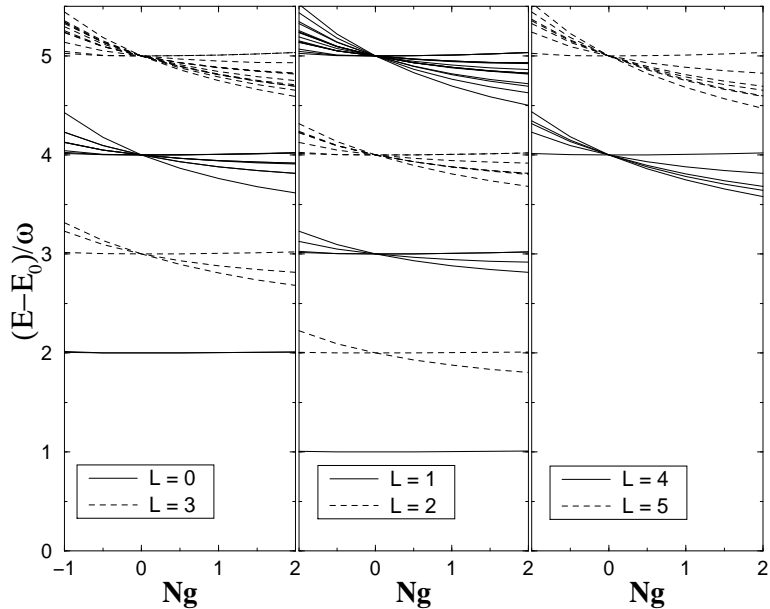,width=10cm,angle=270,height=8cm}}
\end{center}
\caption{\footnotesize Excitation energies $(E-E_0)/\omega$ in a 
two-dimensional system of
20 bosons as function of the interaction strength. L is the angular 
momentum.}
\label{dED2}
\end{figure}
{\em Planar Bose gas.} The lowest energy levels of a planar system with 20 
bosons are given in Fig.\ 7 as function of the dimensionless coupling. 
For each level with non-zero angular momentum $L$, there is a level with the
same energy and angular momentum $-L$. The degeneracy at $g=0$ is partially
lifted for non-zero $g$, 
but again there are levels which are unaffected by the interaction. These
are reached by the two dipole excitations and the breathing mode
excitation from the ground state. The modes also give rise to a set of ladders
above the interaction dependent levels. Since the energies of all three
interaction independent modes are multiples of $\omega$, there is a high
degree of degeneracy even for $g\neq 0$ in this system.

In Fig.\ 6 the corresponding summed density of states $n_{2D}(E)$ 
(left curves) is plotted for 15 bosons with
the coupling running from $Ng=-1$ to 2 in steps of $0.5$. Levels for both
positive and negative angular momentum are included. The step-like form of the
curve for the free case is again smoothed by the interaction. 

For stronger couplings we do not quite reproduce the independence of the 
interaction. This is due to the truncation of the basis of states. However, 
since we know that these levels must approach $(E\!-\!E_0)/\omega =$ 
integer as 
the basis set is increased, the deviation gives a good indication of the 
accuracy of the results. In two dimensions we find that the convergence is
better for those levels involving only dipole excitations than for those which
are reached (partly) by the breathing mode. As discussed above, this is
an expected effect of the existence of the energy cut-off.

\subsection{Ground state density}

\begin{figure}[htb]
\begin{center}
\mbox{\psfig{figure=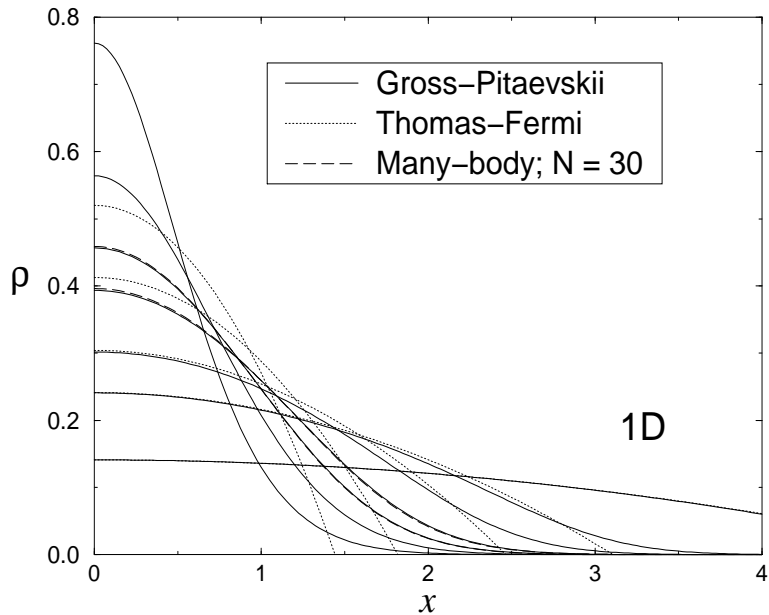,width=10cm,angle=270,height=8cm}}
\end{center}
\caption{\footnotesize Normalized ground state density $\rho$ in one dimension.
The density is plotted as function of the dimensionless
radius $x=r\sqrt{m\omega}$. The solid curves give the density in the 
Gross-Pitaevskii approximation for $Ng=-1,0,1,2,5,10$ and 50.
The dotted curves show the result of the Thomas-Fermi approximation
for repulsive interaction. The density obtained in the exact many-body
calculation with 30 particles is shown as dashed lines for $Ng=1$ and 2.}
\label{GSprofile1}
\end{figure}

The ground state can in the GP-approximation be written as a superposition of 
oscillator functions with even parity or zero angular momentum in one 
and two dimensions respectively. In the many-body description this is no 
longer true, and single particle states with other quantum numbers will 
contribute. We here show how this affects the ground state density.

For large $gN$ the interaction energy 
dominates the kinetic energy, and one may neglect the kinetic term in the 
GP-equation. This is the Thomas-Fermi (TF) approximation \cite{BP}. The 
solution is now simply
\begin{eqnarray}
|\Phi({\bf r})|^2 &=& \frac{1}{2gN}\left(\mu - V_{ex}({\bf r})\right)\theta
\left(\mu - V_{ex}({\bf r})\right)\nonumber\\
&\equiv& \frac{m\omega^2}{4gN}(R^2-r^2)\theta(R^2-r^2).
\end{eqnarray}
This describes a condensate density which vanishes 
outside the radius $R=\sqrt{2\mu/m\omega^2}$. We here compare the normalized 
density $\rho$ in the TF-approximation with the exact solution
of the Gross-Pitaevskii approximation, and also with the full many-body 
calculation. The TF-approximation in three dimensions has been 
studied earlier in \cite{EDCRB}. 

In Fig.\ 8 and Fig.\ 9 we plot the normalized
ground state density $\rho(x)$
as function of the dimensionless radius $x =r\sqrt{m\omega}$ in one and two
dimensions, respectively. The Thomas-Fermi approximation is seen to 
reproduce core density $\rho(x=0)$ for interaction strengths above
$Ng\sim 5$ in one dimension and above $Ng\sim 10$ in two dimensions.
The approximation gets worse as the distance from the core is increased,
where the density has a sharp cut-off in the Thomas-Fermi approximation.
For the lowest values of the interaction strength we compare
the Gross-Pitaevskii approximation to the full many-body calculation with 
$N=30$ and $N=20$ in one and two dimensions, respectively. The GP-approximation
yields quite good results, especially at large distances, but it tends
to overestimate the effect of the interaction. Still, the mean field theory
seems to work better for the density than for the ground state energy. 

\begin{figure}[htb]
\begin{center}
\mbox{\psfig{figure=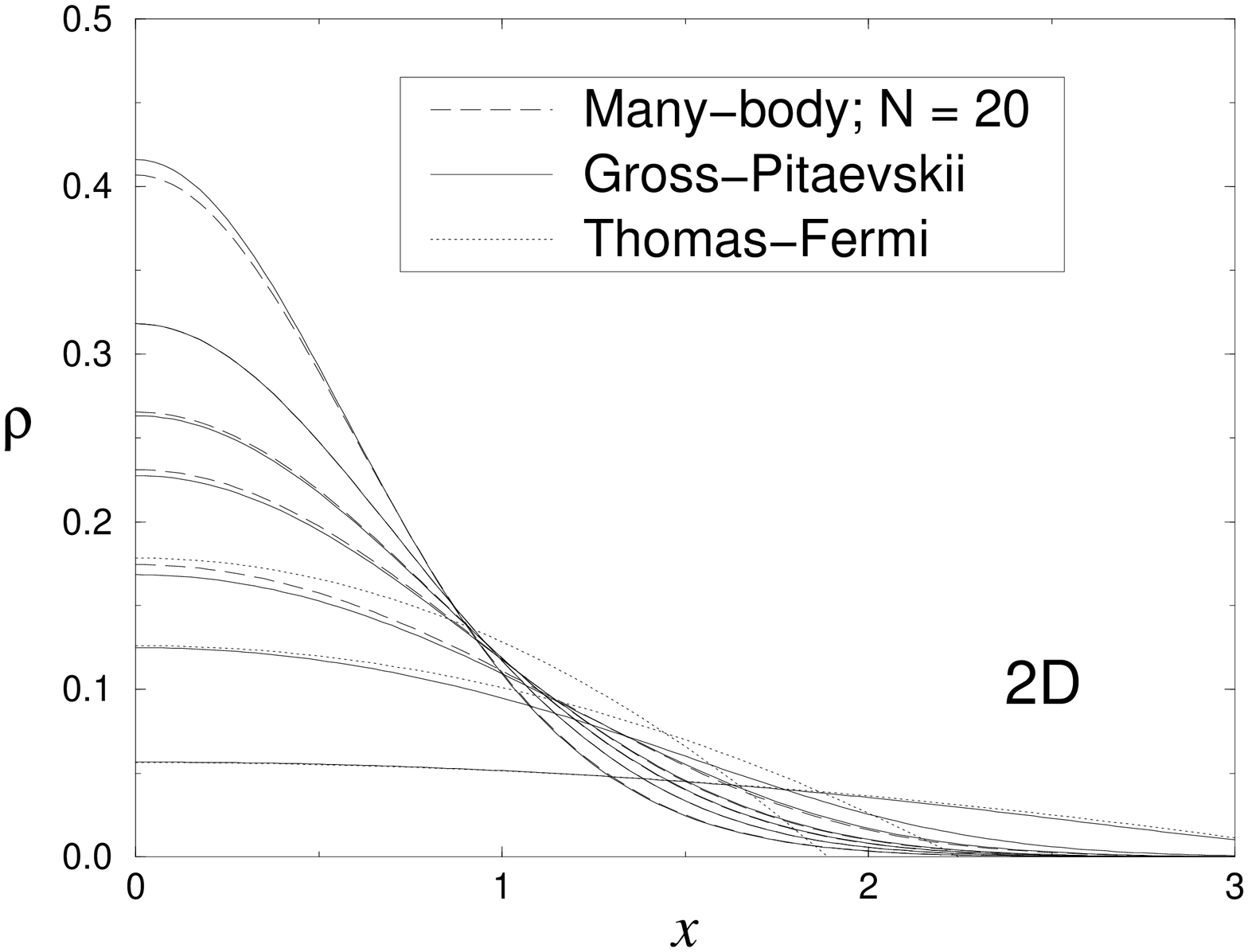,width=10cm,angle=270,height=8cm}}
\end{center}
\caption{\footnotesize Normalized ground state density $\rho$ in two dimensions.
The density is plotted as function of the dimensionless
radius $x=r\sqrt{m\omega}$. The solid curves give the density in the 
Gross-Pitaevskii approximation for $Ng=-1,0,1,2,5,10$ and 50.
The dotted curves show the result of the Thomas-Fermi approximation
for repulsive interaction $Ng = 5, 10$ and 50. The density obtained 
in the exact many-body calculation with 20 particles is shown as
dashed lines for $Ng\le 5$.}
\label{GSprofile2}
\end{figure}

We may again try to compensate
for the deviation by introducing an effective interaction strength, as done
for the energy calculations. The corresponding values of $Ng$
are shown as triangles in Fig.\ 3. The one- and two-dimensional 
results are for $E_{max}/\omega = 38$ and $E_{max}/\omega = 14$, respectively.

The effective couplings are different from those based on the energy 
calculation, and the relative difference increases with $Ng_0$. 
The agreement is however quite good for $Ng_0=0.5$ and 1 in two 
dimensions. But, in general, a simple rescaling of the interaction 
cannot account for all many-body effects in the ground state. 

\subsection{Specific heat}

Having obtained the lowest excitation energies of the system of trapped
bosons, it is straightforward to find the partition function and the 
thermodynamical quantities one may derive thereof. The validity of these
results is limited to low temperatures $T\ll E_{max}$. Just how low the
temperature must be will be clear in the examples which follow.

Let us first consider the one-dimensional system. With 20 particles and a
cut-off at $E_{max}=26\,\omega$, we find the specific heat plotted in the 
left graph of Fig.\ 10.
The coupling ranges from $Ng=0$ to 2 in steps of 0.5, starting from below
at low temperatures. The dotted line is the specific heat for 20 
non-interacting
bosons calculated in the canonical ensemble. We also show the result
of a calculation in the grand canonical ensemble. For particle numbers as low
as this, the two ensembles predict somewhat different specific heats.
The exact curve agrees with the curve for $g=0$ up to a temperature around
$1.5\,\omega$. At higher temperatures the many-body calculation breaks down.
For temperatures where the results are valid we find that the repulsive
interaction yields a higher specific heat than in the case of non-interacting
particles. This happens because the interaction brings most of the energy 
levels
closer to the ground state. The effect of the interaction grows with increasing
temperature in this regime. 

We have also calculated the specific heat for a two-dimensional system of
15 bosons. The cut-off is here $E_{max} = 14\,\omega$, and the basis thus
contains states with angular momentum up to $L=\pm 14$. The specific heat is
shown in the right graph of Fig.\ 10. The calculation now breaks down around 
$T/\omega = 1$. 
For comparison we have included the specific heat of 15 non-interacting bosons
as given in the grand canonical ensemble. Again a repulsive interaction 
increases the specific heat, while it is decreased in the attractive case.

There is no Bose-Einstein condensation for an ideal, {\em homogeneous} gas
in two dimsneions. In the presence of a harmonic trapping potential, 
condensation may however occure \cite{Shevchenko,Mullin}. The condensation 
temperature $T_0 = \hbar\omega\sqrt{N/\zeta(2)}$ persists in the
thermodynamic limit, which is found by taking $N\rightarrow\infty, 
\omega\rightarrow 0$ while keeping the average particle density $\rho\sim 
N\omega^2$ constant. Recently, Mullin \cite{Mullin2} has studied a 
two-dimensional
system with attractive interaction in the thermodynamic limit. He finds that
the normal state has a transition temperature $T<T_0$, but {\em not} to a
Bose condensed phase. This agrees with the earlier findings of Shevchenko
\cite{Shevchenko}.
Keeping, instead $N$ and $\omega$ finite, one should find
a {\em non-zero} temperature at which condensation sets in. Although one
might study this by the method used here, we can not derive any decisive
results as far as the condensation temperature is concerned. For this one 
would need excitation energies going considerably higher than what we have
considered.

\begin{figure}[htb]
\begin{center}
\mbox{\psfig{figure=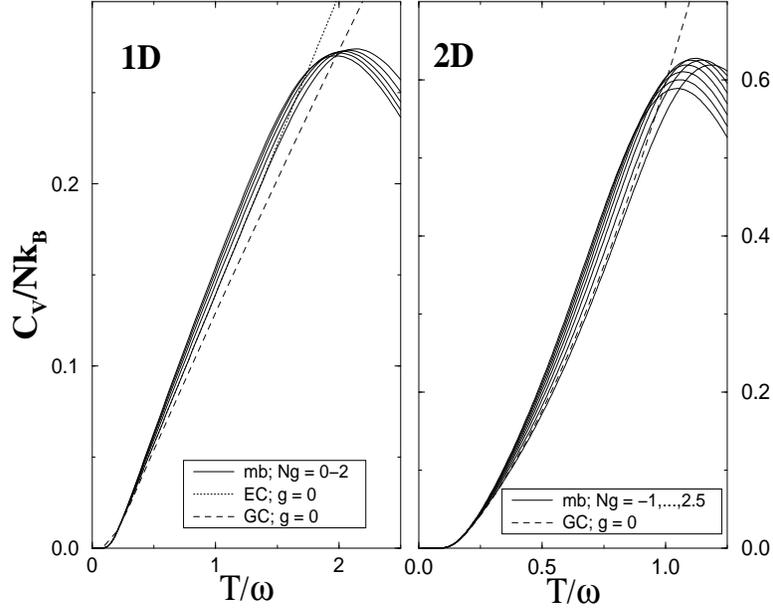,width=10cm,angle=270,height=8cm}}
\end{center}
\caption{\footnotesize Specific heat $C_V$ per particle measured in units of $k_B$
for a one-dimensional (left) and two-dimensional (right) system of 20 and 
15 bosons respectively. 
The specific heat is plotted as function of the reduced temperature $T/\omega$.
In one dimension the interaction strength runs from $Ng = 0$ to 2 and in two
dimensions from -1 to 2.5. $Ng$ increases from below in steps of 0.5.
The many-body results (mb) are
compared to the curve for the ideal gas as given in the grand canonical
ensemble (GC). In one dimension we also show the exact result 
for the ideal gas in the canonical 
ensemble (EC). Notice the breakdown at higher temperatures due to the cutoff
in the energy spectra.
}
\label{CV}
\end{figure}

\section{Discussion and Conclusion}

We have in this paper presented results of calculations in the
configuration interaction approximation on systems
with 10 to 40 interacting bosons confined to harmonic traps in one and
two dimensions. By use of a set of many-body states and decomposition of the
fields in harmonic modes, we have diagonalized the Hamiltonian numerically.
The resulting ground state properties have been compared to the predictions
of the Gross-Pitaevskii equation. 
The largest difference is found for the ground state energy, while for the
ground state density, the mean field predictions are considerably more 
accurate, at least for larger $Ng$.

We have tried to compensate by introducing an effective interaction strength
in the mean-field description. This has been done by computing that 
interaction strength which, when applied in the Gross-Pitaevskii equation 
yields
the same ground state energy and core density as found in the full
many-body calculation. In two dimensions, and for $Ng_0=0.5$ and 1
the effective couplings obtained from matching energy and from matching
core density agree quite well. But, in general, rescaling based on ground 
state energy and core density result in different effective couplings. 
Thus, this simple
rescaling cannot account for all many-body effects in the ground state. 

We have also considered the lowest excitation energies. In both one and 
two dimensions 
we have reproduced the interaction independent levels which must be present
in these systems \cite{FR,PR}. The interaction breaks the
degeneracy in the non-interacting gas completely in one dimension. In two 
dimensions the two dipole modes and the breathing mode assure a high
degree of degeneracy even for non-zero interaction. 
We have calculated the specific
heat based on these excitation energies. For low temperatures, where the
results are valid, the repulsive interaction increases the specific heat,
while it is decreased in the attractive case. 

It would be interesting to compare the obtained excitation energies with 
the results of calculations based on the Bogoliubov or the Hartree-Fock
approximation. One might 
also use the obtained density of states as a basis for calculations
in the microcanonical ensemble. Finally the parameter range might be extended
by choosing a more suitable basis of single particle states, such as the
Hartree-Fock states.

\acknowledgments{The authors would like to thank F. Ravndal for bringing
their attention to this problem and for helpful comments and suggestions.
S. B. Isakov is acknowledged for informing us about reference \cite{FR}.
We thank A. Rosch for pointing out reference \cite{PR} and for useful
comments on the breathing mode in two dimensions.
We also thank B. D. Esry for comments on an early version of the
article.}

\appendix
\section{Effective coupling in lower dimensions}
\label{section:effcoupl}
Inspired by proposed experiments in lower dimensions \cite{KvD,PM}, we will 
here consider the effect of having a strongly anisotropic trapping potential. 
Assuming
$\omega_z\ll\omega_{x,y}$ or $\omega_z\gg\omega_{x,y}$ the system should 
effectively be one- or two-dimensional. When the larger oscillator energy 
dominates all other energies in the system, we may assume that there will be 
no excitations in this direction. This means that the field operator may be 
written as
\begin{eqnarray}
\hat{\Psi}(x,y,z) = \sum_k\phi_0(x)\phi_0(y)\phi_k(z)a_{00k}
\end{eqnarray}
for the one dimensional case, and
\begin{eqnarray}
\hat{\Psi}(x,y,z) = \sum_k\phi_0(z)\phi_k(x,y)a_{0k}
\end{eqnarray}
for the two-dimensional case. We may then integrate over the spatial variables 
corresponding to the stronger oscillation frequencies. This results in a 
rescaling of the coupling constant $g$ in Eq. (\ref{eq:Hi}):
\begin{eqnarray}
g_{2D} = g_{3D}\sqrt{\frac{m\omega_z}{2\pi}}\hspace{0.5cm}
\mbox{and}\hspace{0.5cm}
g_{1D} = g_{3D}\frac{m\omega_{xy}}{2}.
\end{eqnarray}
Measuring energy in units of the remaining oscillator frequency $\omega$ and 
lengths in units of the oscillator length $1/\sqrt{m\omega}$ we obtain the 
dimensionless coupling constants
\begin{eqnarray}
\tilde{g}_{2D} = \sqrt{2\pi}\frac{a}{a_z}\hspace{0.5cm}\mbox{and}\hspace{0.5cm}
\tilde{g}_{1D} = \pi\sqrt{\frac{\omega_{xy}}{\omega_z}}\frac{a}{a_{xy}},
\end{eqnarray}
where $a_i=\sqrt{\hbar/m\omega_i}$ and $a = mg_{3D}/2\pi\hbar^2$. We have here
reintroduced $\hbar$. The effective coupling is seen to increase 
with the strength of the stronger oscillator frequency. This is an effect of 
reducing the length in which the particle density  may extend in the 
dimensions integrated out.

As an example of a possible physical realization, we mention that Ketterle and
van Druten  
\cite{KvD} have suggested an experiment with $\omega_z/\omega_{xy}\sim 
10^{-3}$ and $\omega_{xy}\sim 200 nK$. With a scattering length of typically 
100 Bohr radii, this yields an effective one-dimensional coupling 
$\tilde{g}_{1D} \sim 1$. This corresponds to a strong coupling 
system for any finite number of particles. Other configurations may yield a 
weaker effective coupling.



\end{document}